\documentclass{article}

\def\Lam {\Lambda }
\def\lam {\lambda}

\def \ud {{1 \over 2} }

\def \bea {\begin{equation} }
\def \eea {\end{equation} }

\def \Eslash {E \kern-.5em\slash }
\def \pslash {p \kern-.5em\slash }
\def \kslash {k \kern-.5em\slash }

\def \gsim {\stackrel{>}{\sim}}
\def \lsim {\stackrel{<}{\sim}}
\newcommand{\rpv}{\mbox{$\not \hspace{-0.10cm} R_p$ }}
\newcommand{\rpvi}{\mbox{$\not \hspace{-0.10cm} R_p$}}

\if@twoside\oddsidemargin25mm
\else\oddsidemargin25mm\evensidemargin25mm\marginparwidth25mm\fi
\begin{document}

\large

\title{\bf Status of a Supersymmetric Flavour Violating Solution 
to the Solar Neutrino Puzzle with Three Generations}
\author{H. K. Dreiner$^1$\footnote{e-mail: 
{\tt dreiner@th.physik.uni-bonn.de}}\ , 
G. Moreau$^{1,2}$\footnote{e-mail: 
{\tt moreau@th.physik.uni-bonn.de}} \\ \\
{\it 1: Physikalisches Institut, Nussallee 12 }\\
{\it D-53115 Bonn, Germany } \\ \\
{\it 2: Service de Physique Th\'eorique}\\ 
{CP225, Universit\'e Libre de Bruxelles, Bld. du Triomphe}\\
{\it 1050 Brussels, Belgium }
}
\maketitle

\begin{abstract}
  We present a general study of a three neutrino flavour transition
  model based on the supersymmetric interactions which violate
  R-parity. These interactions induce flavour violating scattering
  reactions between solar matter and neutrinos. The model does not
  contain any vacuum mass or mixing angle for the first generation
  neutrino. Instead, the effective mixing in the first generation is
  induced via the new interactions.  The model provides a natural
  interpretation of the atmospheric neutrino anomaly, and is
  consistent with reactor experiments. We determine all R-parity
  violating couplings which can contribute to the effective neutrino
  oscillations, and summarize the present laboratory bounds.
  Independent of the specific nature of the (supersymmetric) flavour
  violating model, the experimental data on the solar neutrino rates
  and the recoil electron energy spectrum are inconsistent with the
  theoretical predictions. The confidence level of the
  $\chi^2$-analysis ranges between $\sim 10^{-4}$ and $\sim 10^{-3}$.
  The incompatibility, is due to the new SNO results, and excludes the
  present model.  We conclude that a non-vanishing vacuum mixing angle
  for the first generation neutrino is necessary in our model. We
  expect this also to apply to the solutions based on other flavour
  violating interactions having constraints of the same order of
  magnitude.
\end{abstract}

\vskip .5cm


\section{Introduction}
\label{intro}

The most elegant solution of the solar neutrino problem \cite{book} is
the matter-enhanced neutrino oscillation
\cite{Barger80}-\cite{Parke86} (for review articles see
\cite{rev1,rev2}). It is based on the {\it
  Mikheyev-Smirnov-Wolfenstein} (MSW) resonance mechanism \cite{MS,W}
which requires neutrino mixing in vacuum.

An attractive alternative interpretation of the solar neutrino deficit
is the neutrino flavour transition, due to the existence of some
flavour changing (FC) and non-universal flavour diagonal (NFD)
interactions between neutrinos and solar matter
\cite{W}-\cite{Roulet91}.  In such a scenario, neutrino oscillations
are induced by solar matter alone. Thus the neutrinos can even be
massless and no neutrino mixing in vacuum is needed.

The explanation of the solar neutrino puzzle in terms of FC and NFD
scattering of neutrinos on solar matter, has been explored
quantitatively in \cite{Barger91}-\cite{Gago}, and in \cite{Guzzo00}
for the particular case of neutrinos scattering on electrons
characterized by the absence of a resonance. In all these studies, the
predicted neutrino rates are consistent with the results {}from the
solar neutrino experiments (Homestake, GALLEX, GNO, SAGE, SNO,
Kamiokande and Super-Kamiokande). Acceptable fits have also been found
in the framework where both matter-induced neutrino flavour
transitions and neutrino mixing in vacuum exist
\cite{Barger91,Berg97}.

There are also three flavour transition models, solving both the solar
and atmospheric neutrino anomalies and in which FC and NFD
interactions play a fundamental r\^ole. In \cite{Valle01,Valle02} a
model is constructed in which FC and NFD interactions exist between
all neutrino flavours, all the vacuum mixing angles vanish and only
the third neutrino is massive (in vacuum). The authors of \cite{pre}
have considered a scenario where FC and NFD interactions are present
between the first and third neutrino flavours, only the vacuum mixing
angle of $\nu_\mu-\nu_\tau$ sector has a significant value and the
second and third generation neutrinos are massive (in vacuum).  These
scenarios have been chosen so that there is a vacuum contribution to
the mixing in the $\nu_\mu-\nu_\tau$ sector because the atmospheric
neutrino problem cannot be entirely solved by FC and NFD interactions.
The laboratory bounds on the interactions are too severe
\cite{Berg99}. The other main constraint, which these models
satisfied, is the limit on the $\nu_e-\nu_\tau$ mixing angle from
reactor experiments \cite{reac1,reac2,reac3}.

In the present paper, we study the three flavour transition model
where FC and NFD interactions are present between all neutrino
flavours. {\it However}, we restrict the vacuum neutrino masses and a
mixing angle to the second and third generation. We thus address the
question: do we need a vacuum neutrino mixing angle between the first
and second generation to explain the solar neutrino problem or do FC
and NFD interactions suffice?  The model we study is the same model as
in \cite{pre} but with the most general FC and NFD interactions.

We concentrate on the situation where the FC and NFD interactions are
due to supersymmetry \cite{Haber} with broken R-parity
\cite{Salam,Fayet1}, although our analysis is sufficiently general,
that it can be easily applied to any new FC and NFD interaction which
is of the same (small) order of magnitude.  The \rpvi\ superpotential
is written in terms of the left-handed superfields for the leptons
($L$), quarks ($Q$) and Higgs of hypercharge $1/2$ ($H$) and the
right-handed superfields for the charged leptons ($E^c$), up and down
type quarks ($U^c,D^c$),
\begin{eqnarray}
W_{\rpv}=\sum_{i,j,k} \bigg (\ud \lam _{ijk} L_iL_j E^c_k+
\lam^{\prime}_{ijk} L_i Q_j D^c_k+ \ud \lam^{\prime \prime}_{ijk}
U_i^cD_j^cD_k^c
+ \mu_i H L_i \bigg ).
\label{eq:one}
\end{eqnarray}
$i,j,k$ are flavour indices, $\lam_{ijk},\lam'_{ijk},\lam''_{ijk}$ are
dimensionless coupling constants and $\mu_i$ dimension one parameters.
We consider all the relevant \rpv interactions, namely all the
different flavour configurations of both $\lam$ and $\lam'$ couplings, in
contrast with the preliminary study made in \cite{pre} where only a
small set of the $\lam$ couplings were considered.

Based on the most recent Standard Solar Model (SSM): ``BP2000''
\cite{BP2000} (and including the new measurement of $S_{17}(0)$
\cite{S17a,S17b}), we determine whether our generic model remains a
possible interpretation to the solar neutrino anomaly. We have
included the results from the Sudbury Neutrino Observatory (SNO)
\cite{SNOa,SNOb,SNOc,SNOd}. In this sense, our work constitutes an
update of the preliminary analysis in \cite{pre} which was performed
before the results from SNO appeared.

\vspace{-0.4cm}

\section{Description of the Model}

\subsection{Hamiltonian}
\label{Hamilt}

Our model is characterized by the time evolution equation of neutrino
flavour eigenstates,
\begin{eqnarray}
i {d \over dt} \left ( \begin{array}{c}
\nu_e(t) \\ \nu_\mu(t) \\ \nu_\tau(t)
\end{array} \right )
= H \left ( \begin{array}{c}
\nu_e(t) \\ \nu_\mu(t) \\ \nu_\tau(t)
\end{array} \right )\;, 
\label{eq:two}
\end{eqnarray}
where the Hamiltonian has a vacuum contribution as well as a
matter-induced part:
%
\begin{eqnarray}
H &=& E \times {\bf 1}_{3 \times 3}
+ \bigg ({\tilde M^2 \over 2E} \bigg )_{vacuum} 
+ \bigg ({\tilde M^2 \over 2E} \bigg )_{matter}\;, \nonumber \\
 &=& E \times {\bf 1}_{3 \times 3} + \left ( \begin{array}{ccc}
0 & 0 & 0 \\ 
0 & {m_3^2+m_2^2 \over 4E}-{m_3^2-m_2^2 \over 4E}\cos 2\theta^v_{23}  
             &  {m_3^2-m_2^2 \over 4E}\sin 2\theta^v_{23} \\ 
0 & {m_3^2-m_2^2 \over 4E}\sin 2\theta^v_{23} 
             & {m_3^2+m_2^2 \over 4E}+{m_3^2-m_2^2 \over 4E}
\cos 2\theta^v_{23}
\end{array} \right ) \nonumber \\ && \nonumber \\
&&  \quad + \left ( \begin{array}{ccc}
R_{11}+A_1+A_2 & R_{12} & R_{13} \\ 
R_{12} & R_{22}+A_2 & R_{23} \\ 
R_{13} & R_{23} & R_{33}+A_2 
\end{array} \right )\;.
\label{eq:three}
\end{eqnarray}
$E$ is the neutrino energy, ${\bf 1}_{3 \times 3}$ is the $3 \times 3$
identity matrix, $m_2^2$ and $m_3^2$ are, respectively, the second and
third generation neutrino masses (in vacuum).  $\theta^v_{23}$ is the
vacuum mixing angle in the $\nu_\mu-\nu_\tau$ sector, $A_1$ and $A_2$
are the contributions due respectively to the $W^\pm$ and $Z^0$ boson
exchanges.  Note, that we have omitted any first generation
contributions to $({\tilde M}^2/(2E))_{vacuum}$, as discussed in the
introduction.

The $R_{ij}$ terms arise from FC (if $R_{ij} \neq 0 \ [i \neq j]$) and
NFD (if $R_{11} \neq R_{22} \neq R_{33}$) interactions predicted by
new physics: here R-parity violating supersymmetry. $A_1=\sqrt 2 G_F
n_e$ and $A_2=- G_F n_N / \sqrt 2$ and $G_F$ is the {\it Fermi}
coupling constant. $n_e$ is the electron density and $n_N$ the neutron
density in the sun.  Since we concentrate on \rpv interactions, the
$R_{ij}$ terms are given by
\begin{eqnarray}
R_{ij}&=&R_{ij}(\lam)+R_{ij}(\lam')\;, \nonumber \\
R_{ij}(\lam) &=& \bigg ( \sum_{k \neq i,j} 
{\lam_{ik1}\lam_{jk1} \over 4 m^2(\tilde l^\pm_k)} n_e \bigg ) 
+ \bigg ( \sum_{l} 
{\lam_{i1l}\lam_{j1l} \over 4 m^2(\tilde \ell^\pm_l)} n_e \bigg ),
\label{eq:four} \\
R_{ij}(\lam')&=& \sum_{m} \bigg ( 
{\lam'_{im1}\lam'_{jm1} \over 4 m^2(\tilde d_m)} n_d
+ 
{\lam'_{i1m}\lam'_{j1m} \over 4 m^2(\tilde d_m)} n_d \bigg )\;, \nonumber
\end{eqnarray}
where $n_d$ is the down quark density and $m^2(\tilde l^\pm_{k,l})$
($m^2(\tilde d_m)$) is the squared mass of the slepton $\tilde
l^\pm_{k,l}$ (squark $\tilde d_m$) exchanged in the scattering
reaction $\nu_i + e \stackrel{\lam}{\to} \nu_j + e$ ($\nu_i + d
\stackrel{\lam'}{\to} \nu_j + d$). The scattering reaction also fixes
at least one of the generation indices to be 1. The second term of
$R_{ij}(\lam)$ vanishes if $i=1$ or $j=1$, due to the anti-symmetry of
$\lam_{ijk}$ in the first two indices. In the following, we perform a
general analysis with the $R_{ij}$.  Given the approximations
discussed in Section \ref{approx}, this analysis can be applied to any
new FC and NFD interactions.

\subsection{Diagonalization}
\label{diago}

Since the diagonal contributions to the Hamiltonian do not affect 
the flavour transition mechanism, only the following effective mass 
squared matrix,
\begin{eqnarray}
\bigg ({\tilde M^2 \over 2E} \bigg )=H-(E+A_1+A_2)
\times {\bf 1}_{3 \times 3},
\label{eq:five}
\end{eqnarray}
is relevant. It can be diagonalized as,
\begin{eqnarray}
U^\dagger \bigg ({\tilde M^2 \over 2E} \bigg ) U={1 \over 2E}
\left ( \begin{array}{ccc}
\tilde m_1^2 & 0 & 0 \\ 
0 & \tilde m_2^2 & 0 \\ 
0 & 0 & \tilde m_3^2 
\end{array} \right )\;.
\label{eq:six}
\end{eqnarray}
The unitary matrix $U$ is parameterized by the complex phase 
$\delta$ and the mixing angles $\theta_{12}$, $\theta_{13}$ and 
$\theta_{23}$ as,
\begin{displaymath}
U\equiv V_{23} \times V_{13} \times V_{12}\equiv
\left ( \begin{array}{ccc}
1 &    0    & 0 \\ 
0 &  c_{23} & s_{23} \\ 
0 & -s_{23} & c_{23}
\end{array} \right )
\left ( \begin{array}{ccc}
        c_{13}      &  0 & s_{13} e^{-i\delta} \\ 
            0       &  1 & 0 \\ 
-s_{13} e^{i\delta} &  0 & c_{13} 
\end{array} \right )
\left ( \begin{array}{ccc}
 c_{12} & s_{12} & 0 \\ 
-s_{12} & c_{12} & 0 \\ 
   0    &   0    & 1
\end{array} \right )\;,
\end{displaymath}
\begin{eqnarray}
U=
\left ( \begin{array}{ccc}
c_{12}c_{13} & s_{12}c_{13} & s_{13}e^{-i\delta} \\ 
-s_{12}c_{23}-c_{12}s_{23}s_{13}e^{i\delta} 
& c_{12}c_{23}-s_{12}s_{23}s_{13}e^{i\delta}  & s_{23}c_{13} \\ 
 s_{12}s_{23}-c_{12}c_{23}s_{13}e^{i\delta} 
& -c_{12}s_{23}-s_{12}c_{23}s_{13}e^{i\delta} & c_{23}c_{13}
\end{array} \right )\;.
\label{eq:seven}
\end{eqnarray}
Here $s_{ij}=\sin \theta_{ij}$ and $c_{ij}=\cos \theta_{ij}$.  In the
following, we choose the parameters of the Hamiltonian $H$ to be real,
so that $\delta=0$ above. First, we rotate the effective mass squared
matrix by $V_{23}\times V_{13}$ and obtain in the new basis,
\begin{eqnarray}
\bigg ({\tilde M^2 \over 2E} \bigg )=
\left ( \begin{array}{ccc}
R_{11} c_{13}^2-\alpha \sin 2\theta_{13}+\Omega_+ s_{13}^2
& \beta c_{13}
& 0 \\ 
\beta c_{13}
& \Omega_-  
& \beta s_{13} \\ 
0 
& \beta s_{13} 
& R_{11} s_{13}^2+\alpha \sin 2\theta_{13}+\Omega_+ c_{13}^2
\end{array} \right ),
\label{eq:eight}
\end{eqnarray}
with,
\begin{eqnarray}
\alpha\equiv R_{12}s_{23}+R_{13}c_{23},\qquad \beta\equiv R_{12}
c_{23}-R_{13}s_{23},
\label{eq:ten}
\end{eqnarray}
\begin{eqnarray}
\Omega_{\pm}\equiv [{m_3^2+m_2^2 \over 4E}-A_1+{R_{33}+R_{22}\over 2}] 
\pm [{m_3^2-m_2^2 \over 4E}\cos 2(\theta_{23}^v-\theta_{23}) \cr
+{R_{33}-R_{22}\over 2}\cos 2\theta_{23}
+R_{23}\sin 2\theta_{23}].
\label{eq:eleven}
\end{eqnarray}
The mixing angles of Eq.~(\ref{eq:eight})-(\ref{eq:eleven}) are given
in terms of the Hamiltonian parameters (\ref{eq:three})
\begin{eqnarray}
\tan 2\theta_{23}={\sin 2\theta^v_{23} \Delta m_{32}^2/2E+2R_{23}
\over \cos 2\theta^v_{23} \Delta m_{32}^2/2E+R_{33}-R_{22}}\;,
\label{eq:twelve}
\end{eqnarray}
where $\Delta m_{32}^2\equiv m_3^2-m_2^2$, and,
\begin{eqnarray}
\tan 2\theta_{13}={2 \alpha
\over \Omega_+ -R_{11}}\;.
\label{eq:thirteen}
\end{eqnarray}

We show at the end of Section \ref{approx} that, in our framework, the
mixing angle of the $\nu_e-\nu_\tau$ sector has a negligible value
\begin{eqnarray}
\theta_{13} \approx 0\;. 
\label{eq:fourteen}
\end{eqnarray}
Therefore, the upper left $2\times 2$ part of matrix (\ref{eq:eight})
can be readily diagonalized via the mixing angle $\theta_{12}$ defined
by,
\begin{eqnarray}
\tan 2\theta_{12} \approx {2 \beta c_{13}
\over \Omega_- 
-R_{11} c_{13}^2+\alpha \sin 2\theta_{13}-\Omega_+ s_{13}^2}\;.
\label{eq:fifteen}
\end{eqnarray}
Then, the effective mass eigenvalues are,
\begin{eqnarray}
{\tilde m_1^2 \over 2E} &\approx& (R_{11} c_{13}^2-\alpha \sin 2\theta_{13}
+\Omega_+ s_{13}^2) c_{12}^2 -\beta c_{13}\sin 2\theta_{12} 
+\Omega_- s_{12}^2\;,
\label{eq:sixteen} \\
{\tilde m_2^2 \over 2E} &\approx& (R_{11} c_{13}^2-\alpha \sin 2\theta_{13}
+\Omega_+ s_{13}^2) s_{12}^2 +\beta c_{13}\sin 2\theta_{12} 
+\Omega_- c_{12}^2\;,
\label{eq:seventeen} \\
{\tilde m_3^2 \over 2E} &\approx& R_{11} s_{13}^2+\alpha \sin 2\theta_{13}
+\Omega_+ c_{13}^2\;.
\label{eq:eighteen}
\end{eqnarray}

At this stage, an important comment must be made concerning the
$\nu_e-\nu_\mu$ mixing angle $\theta_{12}$. We see from
Eqs.~(\ref{eq:ten}) and (\ref{eq:fifteen}) that $\tan 2\theta_{12}$ is
proportional to the sum of two terms $R_{12}c_{23}$ and
$-R_{13}s_{23}$.  $R_{12}$ constitutes the off diagonal contribution
to the $\nu_e-\nu_\mu$ sector of the Hamiltonian in Eq.~(\ref{eq:three}).
The second term results from a transmission (due to $R_{13}$) of the
mixing in the $\nu_\mu-\nu_\tau$ sector (defined by the angle
$\theta_{23}$) into the $\nu_e-\nu_\mu$ sector.  We conclude that the
$R_{12}$ and $R_{13}$ contributions, induced by \rpv interactions,
generate an effective mixing in the $\nu_e-\nu_\mu$ sector and thus
play a fundamental r\^ole in the neutrino flavour transition
mechanism.

\subsection{Approximations and Constraints from Reactor and
Atmospheric Neutrino Experiments}
\label{approx}

The less stringent bounds on the products of \rpv coupling constants
are typically of order  \cite{Drein}-\cite{perturb}, 
\begin{equation}
\Lam \cdot {\bar\Lam} \lsim  10^{-2}\, \left(\frac{m_{\tilde f}}{100\,
{\rm GeV}}\right)^2\;, \qquad \Lam\;\epsilon \;\
\{\lam,\lam',\lam''\}\;,
\label{eq:bounds}
\end{equation}
where
$m_{\tilde f}$ represents the typical scalar superpartner mass. Hence,
$R_{ij}$ is at most of order ({\it c.f.}\ Section \ref{compar} for a
more precise discussion):
\begin{eqnarray}
R_{ij} \lsim 
 n_x\times2.5\cdot10^{-7}\, {\rm GeV}^{-2}, \quad  
n_x\,\epsilon\,\{n_e,n_d\}.
\label{eq:nineteen} 
\end{eqnarray}
Moreover, inside the sun, $n_e \approx n_d$, the maximal electron
density reached is $n_e^{max} \approx 4.6\cdot10^{11}\, {\rm eV}^3$
\cite{BP2000} and the energy of the produced neutrinos is never higher
than $E=20\,{\rm MeV}$. Therefore,
\begin{eqnarray}
{R_{ij} \over \Delta m^2_{32}/2E} < {\cal O}(10^{-3})\;,
\label{eq:twenty} 
\end{eqnarray}
since $\Delta m^2_{32} \approx 10^{-3}\,{\rm eV}^2$, as we will see in
Eq.~(\ref{consistent}). In the following, we perform an analysis of
the Hamiltonian (\ref{eq:three}) with general entries $R_{ij}$,
however with the restriction (\ref{eq:twenty}), above.  Thus our
analysis applies to any new physics contributing to the $R_{ij}$,
which respects the bound (\ref{eq:twenty}), {\it i.e.}\ with couplings
obeying the bounds (\ref{eq:nineteen}).

We deduce from Eq.~(\ref{eq:fourteen}) that the experimental data on
atmospheric neutrinos \cite{pubatm,lastatm} are consistent with the
$\nu_\mu \to \nu_\tau$ oscillation scenario based on an effective two
flavour mass matrix, for the following values,
\begin{eqnarray}
\Delta \tilde m^2_{32} \equiv \tilde m^2_3 - \tilde m^2_2 \approx 
10^{-3}\, {\rm eV}^2\;,
\label{eq:twentyone} 
\end{eqnarray}
and,
\begin{eqnarray}
\theta_{23} \approx {\pi \over 4}\;.
\label{eq:twentytwo} 
\end{eqnarray}
Eqs.~(\ref{eq:twelve}), (\ref{eq:twenty}) and (\ref{eq:twentytwo})
require for the vacuum mixing angle $\theta^v_{23}$,
\begin{eqnarray}
\theta^v_{23} \approx {\pi \over 4} \approx \theta_{23}\;.
\label{eq:twentythree} 
\end{eqnarray}

We also see from Eq.~(\ref{eq:nineteen}) that the ratio of $R_{ij}$ and $A_1$ 
reaches at most the value, 
\begin{eqnarray}
{R_{ij} \over A_1} \lsim 10^{-2}\;.
\label{eq:twentyfour}
\end{eqnarray}
{}From Eqs.~(\ref{eq:fourteen}), (\ref{eq:fifteen}),
(\ref{eq:twentythree}) and (\ref{eq:twentyfour}), it is clear that the
mixing angle $\theta_{12}$ is given to a good approximation as,
\begin{eqnarray}
\tan 2 \theta_{12} \approx {\sqrt 2 (R_{12}-R_{13})
\over {m_2^2 \over 2E}-A_1}\;.
\label{eq:twentyfive}
\end{eqnarray}
{}From Eq.~(\ref{eq:twentyfive}) we see there is potentially a
resonant point at $\theta_{12} = \pi / 4$. We find that the associated
resonance condition, namely $m_2^2 / 2E = A_1$, can be indeed
satisfied inside the sun. However, for the atmospheric neutrinos,
which have energies of ${\cal O}({\rm GeV})$, this resonance condition
cannot be fulfilled. The relevant parameter values are: $m_2^2 \approx
5 \cdot 10^{-6}\, {\rm eV}^2$ (see Section \ref{fit}) and the electron
density inside the earth $n_e^E\approx (3-6)\,{\rm cm}^{-3}\,{\cal N}_A$,
where ${\cal N}_A$ is {\it Avogadro}'s constant.

{}From Eqs.~(\ref{eq:fourteen}), (\ref{eq:seventeen}),
(\ref{eq:eighteen}), (\ref{eq:twenty}) and (\ref{eq:twentythree}), we
derive the following approximation for the effective mass squared
difference $\Delta \tilde m^2_{32}$ ({\it c.f.}\ 
Eq.~(\ref{eq:twentyone})),
\begin{eqnarray}
\Delta \tilde m^2_{32} \approx \Delta m^2_{32}\;.
\label{eq:twentysix}
\end{eqnarray}
We then deduce from Eqs.~(\ref{eq:twentyone}) and (\ref{eq:twentysix})
that,
\begin{eqnarray}
\Delta m^2_{32} \approx 10^{-3}\, {\rm eV}^2\;.
\label{consistent}
\end{eqnarray}

Besides, Eqs.~(\ref{eq:fourteen}), (\ref{eq:sixteen}),
(\ref{eq:seventeen}), (\ref{eq:twentythree}), (\ref{eq:twentyfour})
and (\ref{eq:twentyfive}) imply that,
\begin{eqnarray}
\Delta \tilde m^2_{21} = \tilde m^2_2 - \tilde m^2_1 
\approx m^2_2 - 2 E A_1.
\label{eq:twentyseven}
\end{eqnarray}

Finally, we justify the approximation of Eq.~(\ref{eq:fourteen}), used
both in the present section and in Section \ref{diago}. First, from
Eqs.~(\ref{eq:twentythree}), (\ref{eq:twentyfour}) and
(\ref{eq:thirteen}), we obtain a good approximation for $\theta_{13}$:
\begin{eqnarray}
\tan 2 \theta_{13} \approx {\sqrt 2 (R_{12}+R_{13})
\over {m_3^2 \over 2E}-A_1}.
\label{eq:twentyeight}
\end{eqnarray}
Now, we see from this result and Eq.~(\ref{eq:twentyfour}) that
$\theta_{13} \approx 0$ outside of the resonance.  The resonance point
associated with $\theta_{13}$ cannot be reached inside the sun,
because the resonance condition $m_3^2 / 2E = A_1$ cannot be
satisfied. Indeed, inside the sun ({\it c.f.}\  Eq.~(\ref{consistent})),
\begin{equation}
\frac{m_3^2} {2E} > \frac{\Delta m_{32}^2}{2E}
\,\gsim\, 2.5 \cdot 10^{-11}\,{\rm eV}\;,\quad 
{\rm and,}\quad A_1 \lsim 7.5 \ 10^{-12}\,{\rm eV}\;.
\end{equation}
The resonance condition $m_3^2 / 2E = A_1$ can also not be fulfilled
for atmospheric neutrinos, given $n_e^E$ above. Moreover, it is clear
from Eq.~(\ref{eq:twentyeight}) that in vacuum, the $\theta_{13}$
mixing angle is exactly equal to zero (like $\theta_{12}$), consistent
with the constraint obtained at reactor experiments (with an effective
two flavour mass matrix): $\sin^2 2 \theta_{13} \lsim
0.1$ for $\Delta m_{31}^2 \gsim 3 \cdot 10^{-3}\, {\rm
  eV}^2$ \cite{reac1,reac2,reac3}.  Therefore, within the present
framework, $\theta_{13}$ approximately vanishes from the point of view
of the reactor, solar and atmospheric neutrino experiments.

\section{Solar Neutrino Rates}

\subsection{Theoretical Transition Probabilities}
\label{theory}

Since to a good approximation $\theta_{13}$ is vanishing, the $\nu_e
\to \nu_\mu$ transitions of solar neutrinos obey a two flavour
transition dynamics described by the upper left $2 \times 2$ block of
the matrix in Eq.~(\ref{eq:eight}). Thus, in the non-adiabatic
transition case, the solar $\nu_e \to \nu_\mu$ transition
probabilities $\bar P_{\nu_e \to \nu_\mu}=1-\bar P_{\nu_e \to \nu_e}$
are defined by \cite{Haxton86,Haxton87},
\begin{equation}
\bar P_{\nu_e \to \nu_e} \simeq {1 \over 2}
\bigg ( 1-(1-2P_x)\cos^2 2\theta^s_{12} \bigg )\,, 
\qquad \bigg [ P_x>{1 \over 2} \bigg ]\;,
\label{eq:twentynine-a}
\end{equation}
for a neutrino produced above the resonance density, and $\bar
P_{\nu_e \to \nu_e} \simeq 1$, for a neutrino produced below the
resonance density. In the adiabatic transition case \cite{Parke86},
\begin{equation}
\bar P_{\nu_e \to \nu_e} \simeq {1 \over 2}
\bigg ( 1+(1-2P_x)\cos 2\theta^p_{12}\cos 2\theta^s_{12} \bigg ) 
\;,\qquad 
\bigg [ P_x<{1 \over 2} \bigg ]\;.
\label{eq:twentynine-b}
\end{equation}
In Eq.~(\ref{eq:twentynine-a}) and Eq.~(\ref{eq:twentynine-b}),
$\theta^p_{12}$ and $\theta^s_{12}$ represent the matter-induced
$\theta_{12}$ mixing angle at the neutrino production point and solar
surface, respectively. $P_x$ is the {\it Landau-Zener-Stueckelberg}
transition probability \cite{LZS},
\begin{eqnarray}
P_x \simeq e^{-{\pi \over 2}\gamma F}.
\label{eq:thirty}
\end{eqnarray}
In this equation, $F \simeq 1-\tan^2 \theta^{vac}_{12}$ \cite{rev2},
and $\theta^{vac}_{12}$ is the $\theta_{12}$ mixing angle in vacuum,
since the electron density inside the sun is approximately given by an
exponential function of the distance $r$ to the solar center
\cite{BP2000}.  From Eq.~(\ref{eq:twentyfive}) we deduce that
$\theta^{vac}_{12} \approx 0$, so that $F \simeq 1$ here.  The
$\gamma$ parameter of Eq.~(\ref{eq:thirty}) is given at the resonance
point,
\begin{eqnarray}
\gamma \equiv \bigg \vert { \Delta \tilde m^2_{21}/2E \over 
2 { d \theta_{12} \over d r } } \bigg \vert_{res}\;.
\label{eq:thirtyone}
\end{eqnarray}
The resonance point is determined by ({\it c.f.}\ 
Eq.~(\ref{eq:twentyfive})),
\begin{eqnarray}
{ m_2^2 \over 2E } = A_1.
\label{eq:thirtytwo}
\end{eqnarray}
Let us derive the expression for $\gamma$ within our model. At the
resonance point, $\theta_{12}=\pi/4$ so that Eqs.~(\ref{eq:sixteen})
and (\ref{eq:seventeen}), together with Eqs.~(\ref{eq:fourteen}) and
(\ref{eq:twentythree}), give, 
\begin{equation}
\left\vert \frac{\Delta \tilde m^2_{21}}{2E}\right\vert_
{res} = \sqrt 2\, \vert R_{12}-R_{13} \vert_{res}\;. 
\end{equation}
{}From Eq.~(\ref{eq:twentyfive}) we see that $\vert 2 \ 
d \theta_{12} / d r \vert_{res} = G_F \vert {d n_e \over d r} /
(R_{12}-R_{13}) \vert_{res}$ and thus, 
\begin{eqnarray}
\gamma = {\sqrt 2 \vert R_{12}-R_{13} \vert^2_{res} \over 
G_F \vert { d n_e \over d r } \vert_{res}}\;.
\label{eq:thirtythree}
\end{eqnarray}

\subsection{Fit with Experimental Results}
\label{fit}

In this subsection, we present the free parameter values giving rise
to the highest degree of consistency between our theoretical
predictions and the experimental solar neutrino rates.  The relevant
free parameters in
Eqs.~(\ref{eq:twentynine-a},\,\ref{eq:twentynine-b}) are $m_2^2$ and
$R$, where
\begin{eqnarray}
R\equiv {\vert R_{12}(\lam) - R_{13}(\lam) \vert 
\over n_e}\;, 
\label{eq:Rlam}
\end{eqnarray}
for $\lam$-\rpv interactions and 
\begin{eqnarray}
R\equiv {\vert R_{12}(\lam') - R_{13}(\lam') \vert 
\over n_d}\;,
\label{eq:Rlamp}
\end{eqnarray}
for $\lam'$-\rpv interactions ({\it c.f.}\ Eq.~(\ref{eq:four})). In
certain specified cases, we also consider $f_B$ as a free parameter:
$f_B$ is defined by $\Phi_{^8\!B}=f_B \times \Phi_{^8\!B}^{SSM}$,
where $\Phi_{^8\!B}$ is the flux of solar electron-neutrinos produced
in the decay $^8\!B \ \to \ ^8\!Be^* + e^+ + \nu_e$ and
$\Phi_{^8\!B}^{SSM}$ is the theoretical prediction within the SSM
\cite{BP2000,S17a,S17b}.  This is equivalent to allowing for an
arbitrary normalization of the $\Phi_{^8\!B}$ flux.

In Table \ref{tab:fit}, we show the values of the free parameters for
which the best fit is obtained between the theoretical solar neutrino
rates defined as,
\begin{eqnarray}
P^{th}_j = 
\sum_i \phi_i \int dE \ dr \ {dg_i(E)\over dE} {dh_i(r)\over dr}
\sigma_j(E) 
\bar P_{\nu_e \to \nu_e}(E,r), 
\label{eq:thirtyfour}
\end{eqnarray}
and the corresponding present experimental solar neutrino rates
$P^{exp}_j$ \cite{SNOa}-\cite{SNOd}, \cite{Homestake}-\cite{SuperK}.
The index $j$ runs over the 5 solar neutrino experiments (we use
the combined result for the GALLEX, GNO and SAGE experiments
\cite{SAGEbis}): $j=$\{Homestake, GALLEX + GNO + SAGE, SNO, Kamiokande,
Super-Kamiokande\}.

Let us briefly describe the parameters entering
Eq.~(\ref{eq:thirtyfour}).  First, $P^{th}_j$ is the theoretical
prediction for the solar neutrino detection rate in experiment $j$, if
neutrinos have a flavour transition average probability $\bar P_{\nu_e
  \to \nu_e}(E,r)$. The $\phi_i$ are the flux of solar
electron-neutrinos, where $i$ labels the source reaction: $i=$\{pp,$^7
$Be, pep, $^{13}$N, $^{15}$O, $^{17}$F, $^8$B, hep\}.  $dg_i(E)/dE$
and $dh_i(r)/dr$ respectively represent the neutrino energy and the
initial radial position distributions (normalized to unity). Both
these fluxes and distributions are taken from the most recent SSM
(BP2000) \cite{BP2000}-\cite{S17b}. Finally, the $\sigma_j(E)$ denote
the neutrino detection cross sections corresponding to the different
solar neutrino experiments \cite{book,XsecCl,XsecGm,XsecCC}.

In the case of the Kamiokande and Super-Kamiokande experiments, the
product $\sigma_j(E)$ $\bar P_{\nu_e \to \nu_e}(E,r)$ in
Eq.~(\ref{eq:thirtyfour}) stands for,
\begin{eqnarray}
\int_{T_{th}}^{T_{max}} \ dT \ \int_0^{\infty} \ dT' \ 
\epsilon (T') \ \rho (T,T') \ 
\bigg [ 
{d\sigma_j^{\nu_e}(E,T') \over dT'}\bar P_{\nu_e \to \nu_e}(E,r)
 \cr
+{d\sigma_j^{\nu_\mu}(E,T') \over dT'}
\bigg ( 1 - \bar P_{\nu_e \to \nu_e}(E,r) \bigg ) 
\bigg ]\;,
\label{eq:thirtyfive}
\end{eqnarray}
where $T'$ is the true energy of the recoil electron, $T$ its value
measured experimentally, $T_{th}$ the energy threshold of the
experiment, $T_{max}$ the maximal energy of the experiment, $\epsilon
(T')$ the detection efficiency, $\rho (T,T')$ the energy resolution
function and $d\sigma_j^{\nu_e}(E,T')/dT'$
($d\sigma_j^{\nu_\mu}(E,T')/dT'$) the differential cross section for
the solar electron-neutrino (muon-neutrino) elastic scattering
reaction on electrons \cite{book}.  The resolution function of
Eq.~(\ref{eq:thirtyfive}) can be satisfactorily represented by the
following Gaussian function \cite{Lisi97},
\begin{eqnarray}
\rho (T,T')= { e^{{-(T-T')^2 \over 2 \Delta^2(T')}}
\over \sqrt{2 \pi \Delta^2(T')}}\;.
\label{eq:gaussian}
\end{eqnarray}
The standard deviation of the energy measurement is approximately
given by,
\begin{eqnarray}
{\Delta(T') \over T'} \approx a \ \bigg ( 
{10 MeV \over T'} \bigg )^{1/2}.
\label{eq:deviation}
\end{eqnarray}
We take $T_{th}=7.5\,{\rm MeV}$ \cite{Kamiokande}, $T_{max}=20\,{\rm
  MeV}$ \cite{Kamiokande}, $\epsilon \approx 0.7$ \cite{book} and
$a=0.143$ \cite{Kamiokande,Eres1,Eres2,Eres3}, for the Kamiokande
detector, and $T_{th}=5\,{\rm MeV}$ \cite{SuperK}, $T_{max}=20\,{\rm
  MeV}$ \cite{SuperK}, $\epsilon \approx 0.7$ \cite{SuperK,SuperK98}
and $a=0.153$ \cite{Nakahata99,Faid97}, for the Super-Kamiokande one.

For the SNO experiment, the product $\sigma_j(E) \bar P_{\nu_e \to
  \nu_e}(E,r)$ in Eq.~(\ref{eq:thirtyfour}) stands for,
\begin{eqnarray}
\int_{T^{SNO}_{th}}^{T^{SNO}_{max}} \ dT \ \int_0^{\infty} \ dT' \ 
\rho (T,T') \ 
{d\sigma_j^{CC}(E,T') \over dT'}\bar P_{\nu_e \to \nu_e}(E,r)\;.
\label{eq:thirtyfiveSNO}
\end{eqnarray}
$d\sigma_j^{CC}(E,T')/dT'$ is the differential cross section of the
charged current reaction between the solar electron-neutrinos and
deuterium: $\nu_e+D \to p+p+e^-$ \cite{XsecCC}. The energy resolution
function $\rho (T,T')$ of Eq.~(\ref{eq:thirtyfiveSNO}) is defined as
before by Eq.~(\ref{eq:gaussian}), but with the following standard
deviation $\Delta(T')$ \cite{SNOa},
\begin{eqnarray}
\Delta(T') = \bigg [ - 0.4620 + 0.5470\, \sqrt{T'/1 \,{\rm MeV}} 
+ 0.008722\, (T'/1 \,{\rm MeV}) \bigg ]\, {\rm MeV}.
\label{eq:deviationSNO}
\end{eqnarray}
In the present analysis, we use $T^{SNO}_{th}=6.75\,{\rm MeV}+m_e$
\cite{SNOa}, $m_e$ is the electron mass, and $T^{SNO}_{max}=13\,{\rm
  MeV}+m_e$ \cite{SNOa}.  

The points of parameter space corresponding to the best fit between
the theoretical and experimental solar neutrino rates have been
obtained by performing a $\chi^2$-analysis.  Following the analyses of
\cite{Smirnov98,Garay99}, we define our $\chi^2$-function as,
\begin{eqnarray}
\chi^2_R = \sum_{j_1,j_2} 
(P^{th}_{j_1}-P^{exp}_{j_1})
[\sigma^2_R]^{-1}_{j_1j_2}
(P^{th}_{j_2}-P^{exp}_{j_2}),
\label{eq:thirtysix}
\end{eqnarray}
where $j_1$ and $j_2$ run over the solar neutrino experiments and
$[\sigma^2_R]_{j_1j_2}$ is the squared error matrix containing both
the experimental (systematic and sta\-tis\-tical)
\cite{SNOa,SNOb,Homestake}, \cite{SAGEbis}-\cite{SuperK} and theoretical
errors (computed according to the analyses of \cite{Fogli95,Guzzo98})
on the solar neutrino rates.  The theoretical errors depend on the
uncertainties of both the neutrino fluxes \cite{BP2000,S17b},
\cite{BU}-\cite{BP1998} and the detection cross sections
\cite{XsecCl}-\cite{XsecCC}, \cite{XsecCCpre}.

In Table \ref{tab:fit}, our results concerning the fits of solar
neutrino rates are given in two different contexts: before (case (1))
and after the publication of experimental data from SNO
\cite{SNOa}-\cite{SNOd} (case (2)).  In the ``post-SNO'' context, we
include into our $\chi^2$-analysis the rate of solar
electron-neutrinos detected at the SNO experiment via the charged
current reaction on deuterium \cite{SNOa}. Furthermore, within this
context, we fix the normalization factor $f_B$ to unity, following the
study of \cite{Gago}. The reason is that the value of the neutrino
flux $\Phi_{^8\!B}$ measured at the SNO experiment (with the flux of
neutrinos detected through the neutral current reaction on deuterium)
\cite{SNOa,SNOb,SNOd} is in agreement with its prediction in the SSM
\cite{BP2000,S17a,S17b}: $\Phi_{^8\!B}^{SSM}$.  Finally, let us note
that within this post-SNO context, we do not implement into the
$\chi^2$-analysis the rate of solar neutrinos detected at the SNO
experiment via the elastic scattering reaction on electrons
\cite{SNOa,SNOb,SNOd}. This is justified by the fact that this rate is
consistent with the rates of neutrinos detected at the Kamiokande and
Super-Kamiokande experiments via the same scattering reaction
\cite{Kamiokande,SuperK}, which are already taken into account in our
analysis.

\begin{table}[t!]
\begin{center}
\begin{tabular}{|c|c|c|c|c|c|c|c|c|c|}
\hline
Model & $m_2^2$ & $R$ & & $\chi^2$ 
& H & G+G+S & SNO & K & SK \\ 
(fit) & $(10^{-5}$ & $(10^{-24}$ & $f_B/{\bar\alpha}$ & $(DOF; C.L.)$
& $(0.297$ & $(0.557$ 
& $(0.295$ & $(0.472$ & $(0.391$ \\
& $eV^2)$ & $ eV^{-2})$ & & & $\pm 0.026)$ & $\pm 0.039)$ 
& $\pm 0.023)$ & $\pm 0.064)$ & $\pm 0.014)$ \\
\hline
$\lam$ &
0.545 & 1.319 & $f_B=$ & 0.19
& 0.295 & 0.574 & $\times$ & 0.476 & 0.390 \\
(1) & & & 1.087 & (1; $66.3 \%$)
& & & & & \\
\hline
$\lam$ &       
0.625 & 1.055 & & 6.25
& 0.244 & 0.595 & 0.316 & 0.402 & 0.369 \\
(2) & & & & (3; $10.0 \%$)
& & & & & \\
\hline
$\lam$ &       
0.648 & 0.561 & ${\bar\alpha}=$ & 46.60
& 0.560 & 0.645 & 0.704 & 0.738 & 0.713 \\
(3) & & & 0.624 & (21; $0.11\%$)
& & & & & \\
\hline
$\lam'$ &      
0.534 & 1.112 & $f_B=$ & 0.18
& 0.303 & 0.562 & $\times$ & 0.491 & 0.389 \\
(1) & & & 1.115 & (1; $67.1 \%$)
& & & & & \\
\hline
$\lam'$ &      
0.625 & 0.814 & & 6.23
& 0.244 & 0.595 & 0.316 & 0.402 & 0.369 \\
(2) & & & & (3; $10.1 \%$)
& & & & & \\
\hline
$\lam'$ &      
0.386 & 0.257 & ${\bar\alpha}=$ & 51.73
& 0.865 & 0.629 & 0.997 & 0.992 & 0.972 \\
(3) & & & 0.459 & (21; $0.021\%$)
& & & & & \\
\hline
\end{tabular}
\caption{ The values of the parameters $m_2^2$, $R$, $f_B$ and 
  ${\bar\alpha}$ (see text) associated with the best fits between the
  solar neutrino experimental data and the theoretical prediction.
  They are given in the 3 cases we consider: (1) the solar neutrino
  rates except those of the SNO experiment, (2) all the solar neutrino
  rates, including the rates detected at the SNO experiment via the
  charged current reaction, and (3) all the solar neutrino rates
  together with the recoil electron energy spectrum obtained at the
  Super-Kamiokande experiment. Both the models based on the \rpv
  interactions of type $\lam$ and $\lam'$ are considered. We show the
  values, computed with the obtained best-fit parameters, of the
  relevant $\chi^2$-function (with associated number of degrees of
  freedom denoted as DOF and confidence level denoted as C.L.), namely
  either $\chi^2_R$ (for cases 1 and 2) or $\chi^2_S$ (for case 3). We
  also indicate, for the best-fit parameters, the theoretical solar
  neutrino rates $P^{th}_j$ (see Eq.~(\ref{eq:thirtyfour})) normalized
  to their value expected in the SSM BP2000 \cite{BP2000,S17a,S17b},
  at the considered experiments: Homestake (H), GALLEX+GNO+SAGE
  (G+G+S), SNO, Kamiokande (K) and Super-Kamiokande (SK). The
  corresponding present experimental solar neutrino rates $P^{exp}_j$
  (together with their experimental uncertainty)
  \cite{SNOa}-\cite{SNOd}, \cite{Homestake}-\cite{SuperK} are written
  in parentheses. Finally, we note that when the value of $f_B$ or
  ${\bar\alpha}$ (which are presented in the same column) is not
  given, it means that it is fixed to unity and/or it is not
  relevant.}
\label{tab:fit}
\end{center}
\end{table}

In Table \ref{tab:fit}, we also present the values of the parameters
corresponding to the best fit of all 5 solar neutrino rates {\it and}
the recoil electron energy spectrum (case (3)). This fit takes into
account the consistency between the recoil electron energy spectrum
obtained with the Super-Kamiokande detector from 1258 days of data
\cite{SuperK} and its theoretical prediction. These results have also
been derived through the $\chi^2$-method with the $\chi^2$-function
\cite{Smirnov98,Garay99},
\begin{eqnarray}
\chi^2_S = \chi^2_R + \sum_{j_1,j_2} 
({\bar\alpha} S^{th}_{j_1}-S^{exp}_{j_1})
[\sigma^2_S]^{-1}_{j_1j_2}
({\bar\alpha} S^{th}_{j_2}-S^{exp}_{j_2})\;.
\label{eq:chi2S}
\end{eqnarray}
Here the $S^{exp}_j \ [j=1,...,19]$ are the solar neutrino rates
observed in the Super-Kamiokande experiment corresponding to the 19
bins of a given recoil electron energy range \cite{SuperK} and the
$S^{th}_j \ [j=1,...,19]$ are their predicted values. The $S^{th}_j$
are calculated with Eqs.~(\ref{eq:thirtyfour})--(\ref{eq:deviation})
(applied to the Super-Kamiokande experiment) by integrating $T$ over
the relevant 19 energy intervals \cite{SuperK}. The matrix
$[\sigma^2_S]_{j_1j_2}$ entering Eq.~(\ref{eq:chi2S}) is equal to
\cite{Berg00},
\begin{eqnarray}
[\sigma^2_S]_{j_1j_2}=\delta_{j_1j_2}
([\sigma^2_{stat}]_{j_1j_2}+[\sigma^2_{uncorr}]_{j_1j_2})
+[\sigma^2_{corr}]_{j_1j_2}+[\sigma^2_{th}]_{j_1j_2}\;.
\label{eq:sigma2S}
\end{eqnarray}
The $[\sigma^2_{stat}]_{j_1j_2}$ represent the squared experimental
sta\-tis\-tical errors \cite{SuperK}, $[\sigma^2_{corr}]_{j_1j_2}$ and
$[\sigma^2_{uncorr}]_{j_1j_2}$ are respectively the correlated and
uncorrelated contributions to the squared experimental systematic
errors \cite{SuperK} and $[\sigma^2_{th}]_{j_1j_2}$ are the squared
theoretical uncertainties \cite{BP2000,S17b},
\cite{Fogli95}-\cite{BP1998}.  Finally, the overall normalization
factor ${\bar\alpha}$ introduced in Eq.~(\ref{eq:chi2S}) has been taken as
an additional free parameter in the fit, in order to avoid
double-counting with the data {}from the Super-Kamiokande experiment
on the total event rate which is already included in $\chi^2_R$.

Let us comment on the results of the $\chi^2$-analyses given in Table
\ref{tab:fit}. In case one considers the solar neutrino rates
excluding the SNO experiment, case (1), one obtains acceptable fits
between the theoretical and the experimental rates, namely fits at
$66.3 \%$ and $67.1 \%$ of Confidence Level (C.L.) in the scenarios
containing $\lam$ and $\lam'$ interactions, respectively.  When adding
the SNO results to the analysis, the fit grows significantly worse:
the theoretical predictions are now compatible with the experimental
data only at the $10.0 \%$ C.L. and $10.1 \%$ C.L., respectively.
Finally, considering all the solar neutrino rates (including the SNO
data) as well as the recoil electron energy spectrum, case (3), the
obtained fits have a confidence level of $1.1 \times 10^{-3}$ C.L. and
$2.1 \times 10^{-4}$ C.L., respectively: the theoretical and
experimental results are not compatible in this case. This
incompatibility is due to the SNO results and not the recoil electron energy
spectrum. In \cite{pre} the authors obtained an acceptable fit, without
the SNO data but including the recoil electron energy spectrum.

Furthermore, we observe in Table \ref{tab:fit} that the values of the
quantity $R$, corresponding to the best fit, are smaller in case the
solar neutrino flavour transitions are induced by \rpv interactions of
the type $\lam'$ instead of $\lam$.  The reason is that the quantity
({\it c.f.}\ Eq.~(\ref{eq:Rlam})),
\begin{equation}
X_{\lam}=\vert R_{12}-R_{13} \vert=\vert R_{12}(\lam)-R_{13}
(\lam) \vert=R n_e\;,
\end{equation}  
entering the transition probability $\bar P_{\nu_e \to \nu_e}$ (see
Section \ref{theory}), in case neutrino flavour transitions are due to
$\lam$ couplings, is replaced by ({\it c.f.}\ Eq.~(\ref{eq:Rlamp})),
\begin{equation}
X_{\lam'}=\vert R_{12}-R_{13} \vert=\vert
R_{12}(\lam')-R_{13}(\lam') \vert=R n_d\;.
\end{equation} 
The decisive difference is that $0.50 \ n_d \lsim n_e \lsim 0.78 \ 
n_d$ inside the sun \cite{Berg00,BP2000}. In other words, a
modification of the matter-induced transition probability $\bar
P_{\nu_e \to \nu_e}$ due to an increase of the solar matter density
can be compensated by a decrease of effective neutrino \rpv
interaction strength (quantified here by $R$). Let us note that this
compensation is not exact, as the solar matter density depends on the
radial position, while this is not true for the parameter $R$.  This
is why, in Table \ref{tab:fit}, the best-fit values of the other free
parameters, $m_2^2$ and $f_B$ (or ${\bar\alpha}$), are also different
in the two situations where flavour transitions are due to $\lam$ and
$\lam'$ couplings.

Let us make a brief final remark: we have found that, in the scenario
where solar neutrino flavour transitions are simultaneously induced by
\rpv interactions of type $\lam$ and $\lam'$, the fits between
observed and predicted solar neutrino rates and recoil electron energy
spectrum are not significantly improved compared to the fits for which
the results are shown in Table \ref{tab:fit}.

\subsection{Comparison of the Results with the Bounds on the
\rpv Coupling Constants}
\label{compar}

In this section, we check that the relevant present limits on \rpv
coupling constants are compatible with the values of $R$ giving rise
to the best fits of solar neutrino rates (see previous section).  This
check is also a self-consistency check. It allows us to verify that
the typical bound on the relevant \rpv coupling constants
(\ref{eq:nineteen}), on which the approximations (see Section
\ref{approx}) used in the mass matrix diagonalization are based, are
correct.

$\bullet$ {\bf $\lam$ coupling constants:} For $\lam$-couplings
$R$ reads explicitly ({\it c.f.}\ Eq.~(\ref{eq:four})),
\begin{eqnarray}
R= \bigg \vert {\lam_{131}\lam_{231} \over 4 m^2(\tilde \tau^\pm)}
-{\lam_{121}\lam_{321} \over 4 m^2(\tilde \mu^\pm)} \bigg \vert.
\label{eq:thirtyseven}
\end{eqnarray}
The largest best fit value for $R$ in Table \ref{tab:fit} is $1.319 \ 
10^{-24} eV^{-2}$.  The bounds on the R-parity violating couplings are
usually given as bounds on a single coupling or on a product of two
couplings. This is based on the assumption that the other couplings
vanish\footnote{The assumption that a single or a pair of \rpv
  coupling constants are dominant is often adopted in the literature,
  as a simplification.  This hypothesis is justified by the analogy
  between structures of \rpv and Higgs Yukawa interactions
  \cite{dreiner}. The latter exhibit a strong hierarchy in flavour
  space.}.  We adopt this approximation in the following.

No bound exists on the \rpv coupling constant product
$\lam_{121}\lam_{321}$. The present experimental constraints at
$2\sigma$ on the single \rpv coupling constants $\lam_{121}$ and
$\lam_{321}$ are given respectively by
\cite{Drein,actualize,perturb,Barger89},
\begin{eqnarray}
\lam_{121}<0.049 \ \bigg ( {m(\tilde e_R^\pm) \over 100\, 
{\rm GeV}} \bigg ),
\qquad
\lam_{321}<0.07 \ \bigg ( {m(\tilde e_R^\pm) \over 100\, 
{\rm GeV}} \bigg )\;.
\label{eq:bound321}
\end{eqnarray}
Note that a different slepton mass appears in $R$ and in the bound
above.  If we assume that $m({\tilde\mu})=m(\tilde e_R)$, we obtain
from the second term in Eq.~(\ref{eq:thirtyseven}) that
$R<9\cdot10^{-26}\,{\rm eV}^{-2}$ and this model irrelevant for FC and
NFD interactions in the sun. If however $m({\tilde\mu})=m(\tilde
e_R)/4$, which could very well be the case with non-universal boundary
conditions at the unification scale,
\begin{equation}
R<1.372 \cdot 10^{-24}\, {\rm eV}^{-2},
\end{equation}  
and the $\lam_{121}\lam_{321}$ interactions can reach the best fit
values for $R$.

The strongest bound on the Yukawa coupling constant product $\lam_
{131}\lam_{231}$ is \cite{Roy96},
\begin{eqnarray}
\lam_{131}\lam_{231}<6.6 \ 10^{-7} \ \bigg ( {m(\tilde \nu_{\tau L}) 
\over 100\, {\rm GeV}} \bigg )^2.
\label{eq:bound131and231}
\end{eqnarray}
In this case we would need to require $m({\tilde \nu}_{\tau L})\approx
300\times m({\tilde\tau})$ in order to obtain a sufficiently large
$R$.  For $m({\tilde\tau})\gsim100\,{\rm GeV}$ (the LEP bound
\cite{Mlim}), this is not compatible with a supersymmetric solution to
the hierarchy problem. We conclude that these interactions can not
significantly contribute to FC and NFD interactions in the sun.

In summary, only when $R$ is given by $R=\vert \lam_{121}\lam_{321}
\vert / 4 m^2(\tilde \mu^\pm)$, the present constraints on relevant
\rpv interactions allow $R$ to take the relevant value of $1.319 \ 
10^{-24} eV^{-2}$.

$\bullet$ {\bf $\lam'$ coupling constants:} Let us now consider the
model (not studied in paper \cite{pre}) in which neutrino flavour
transitions are due to $\lam'$ \rpv-interactions via solar down
quarks. 
\begin{eqnarray}
R=\bigg \vert \sum_{m} \bigg ( 
{\lam'_{1m1}\lam'_{2m1} \over 4 m^2(\tilde d_m)} + 
{\lam'_{11m}\lam'_{21m} \over 4 m^2(\tilde d_m)}
- 
{\lam'_{1m1}\lam'_{3m1} \over 4 m^2(\tilde d_m)} -
{\lam'_{11m}\lam'_{31m} \over 4 m^2(\tilde d_m)} \bigg ) 
\bigg \vert.
\label{eq:thirtyeight}
\end{eqnarray}
Based on the present constraints on \rpv interactions, we determine
whether $R$ can reach $1.11 \ 10^{-24} eV^{-2}$, which is the maximal
value appearing in Table \ref{tab:fit}.  As before, we consider
separately each term of Eq.~(\ref{eq:thirtyeight}), supposing that it
dominates over all the others.

The product $\lam'_{121}\lam'_{321}$ (third term in $R$ with $m=2$) is
only constrained by the present individual experimental limits at
$2\sigma$ on $\lam'_{121}$ and $\lam'_{321}$
\cite{actualize,perturb,Barger89},
\begin{eqnarray}
\lam'_{121}<0.043 \ \bigg ( {m(\tilde d_R) \over 100\, {\rm GeV}} 
\bigg )\;, \qquad
\lam'_{321}<0.52 \ \bigg ( {m(\tilde d_R) \over 100\, {\rm GeV}} \bigg )\;.
\label{eq:boundp321}
\end{eqnarray}
Again a different squark mass appears in $R$, $m({\tilde s})$, than in
the bounds, $m({\tilde d}_R)$. If we assume they are equal, then
\begin{equation}
R<5.6\cdot10^{-25}\,{\rm eV}^{-2}\;.
\end{equation}
This is just below the maximal best-fit value. For a moderate
non-degeneracy of $m(\tilde d_R)=1.5\times m(\tilde s)$, the maximal
best-fit value can be obtained.

The Yukawa coupling constant product $\lam'_{131}\lam'_ {331}$ (third
term in $R$, with $m=3$) is only constrained by the present individual
experimental bounds on $\lam'_{131}$ (at $3\sigma$)
\cite{perturb,Barger89} and on $\lam'_{331}$ (at $2\sigma$)
\cite{Bhatt,perturb,Ellis95}:
\begin{eqnarray}
\lam'_{131}<0.019 \ \bigg ( {m(\tilde t_L) \over 100\,{\rm GeV}} \bigg )\;,\qquad
\lam'_{331}<0.45 \quad {\rm at}\; m(\tilde d) = 100\,{\rm GeV} \;.
\label{eq:boundp131}
\end{eqnarray}
Assuming all squark masses are equal
\begin{equation}
R<2.1\cdot10^{-25}\,{\rm eV}^{-2}\;,
\end{equation}
which is below the best-fit values in Table \ref{tab:fit}. Given a
non-degeneracy of $m(\tilde d)=m(\tilde t_L)=2.3\times m(\tilde b)$,
the fit values for $R$ can be reached.

For $m=1$ we have for the last two terms the coupling $\lam'_{111}
$, which has the strict bound at $2\sigma$
\cite{perturb,Mohapatra86,Hirsch95},
\begin{eqnarray}
\lam'_{111}<5.2 \ 10^{-4} \ 
\bigg ( {m(\tilde e) \over 100\,{\rm GeV}} \bigg )^2
\bigg ( {m(\tilde \chi^0) \over 100\,{\rm  GeV}} \bigg )^{1/2}\;.
\label{eq:boundp111}
\end{eqnarray}
In these cases, we can therefore not reach the fit values for $R$, unless
we require an extreme hierarchy in superpartner masses.

The present experimental bounds at $2\sigma$ on $\lam'_{312}$ and
$\lam'_{112}$ \cite{actualize,perturb,Barger89} (fourth term in $R$
with $m=2$) are,
\begin{eqnarray}
\lam'_{312}<0.11 \ \bigg ( {m(\tilde s_R) \over 100 GeV} \bigg )\;,
\qquad
\lam'_{112}<0.021 \ \bigg ( {m(\tilde s_R) \over 100 GeV} \bigg )\;.
\label{eq:boundp112}
\end{eqnarray}
In this case we always have the same squark mass and we can deduce a rigorous
bound 
\begin{equation}
R<5.8\cdot10^{-26}\,{\rm eV}^{-2}\;,
\end{equation}
which excludes any relevant contribution to FC and NFD interactions in
the sun.

In the same way, the present experimental limits at $2\sigma$ on
$\lam'_{313}$ and $\lam'_{113}$ \cite{actualize,perturb,Barger89},
\begin{eqnarray}
\lam'_{313}<0.11 \ \bigg ( {m(\tilde b_R) \over 100 GeV} \bigg )\;,
\quad 
\lam'_{113}<0.021 \ \bigg ( {m(\tilde b_R) \over 100 GeV} \bigg )\;,
\label{eq:boundp113}
\end{eqnarray}
result in $R<5.8\cdot10^{-26}\,{\rm eV}^{-2}\;$, precluding any
significant contribution.

The three products of Yukawa coupling constants
$\lam'_{11m}\lam'_{21m} \;, m=1,2,3\,,$ (second term in $R$) suffer
the following strong constraint \cite{Kim97},
\begin{eqnarray}
\lam'_{11k}\lam'_{21k}<1.7 \cdot 10^{-7} \ \bigg ( {m(\tilde d_{k R}) 
\over 100 GeV} \bigg )^2\;.
\label{eq:boundp11kandp21k}
\end{eqnarray}
Hence, $R<4.25\cdot10^{-30} {\rm eV}^{-2}$ and no interesting
solutions exist.

The products $\lam'_{1m1}\lam'_{2m1} \;, m=1,2,3$ (first term in $R$)
also have a stringent limit \cite{Kim97}:
\begin{eqnarray}
\lam'_{1j1}\lam'_{2j1}<1.6 \ 10^{-7} \ \bigg ( {m(\tilde u_{j L}) 
\over 100 GeV} \bigg )^2\;,
\label{eq:boundp1j1andp2j1}
\end{eqnarray}
precluding any significant contributions to the FC and NFD
interactions in the sun.

In summary, if the two coupling constants $\lam'_{121}$ and
$\lam'_{321}$, or $\lam'_{131}$ and $\lam'_{331}$, are dominant so
that the quantity $R$ is given by $R=\vert \lam'_{121}\lam'_{321}
\vert / 4 m^2(\tilde s)$, or $R=\vert \lam'_{131}\lam'_{331} \vert / 4
m^2(\tilde b)$, respectively, the present bounds on relevant \rpv
coupling constants allow $R$ to reach the relevant value: $1.112 \ 
10^{-24} eV^{-2}$.

\section{Conclusions}

We have studied a three neutrino flavour transition model (see Section
\ref{Hamilt}), based on the supersymmetric $\not\!\!R_p-$interactions
between neutrinos and solar matter (see Section \ref{diago}).  We have
not included a vacuum mass or any vacuum mixing angles for the first
generation neutrino. Instead the effective mixing is induced via
the flavour violating interactions.

Within this scenario, we have found that, among all the (\rpv)
contributions $R_{ij}$ to the matter-induced part of the effective
mass matrix in Eq.(\ref{eq:three}), only $R_{12}$ and $R_{13}$ play a
r\^ole (see Section \ref{approx}).  Then we have shown, from a
systematic study of the constraints on the \rpv couplings, that only
the products $\lam_{121}\lam_ {321}$, $\lam'_{121}\lam'_{321}$ and
$\lam'_{131} \lam'_{331}$, which enter $R_{13}$, can result in a
sufficiently large value of $R$ in agreement with the best fit
obtained for the solar neutrino data (see Section \ref{compar}).

However, whatever the flavour configurations of the \rpv contributions
are, we have found that the best fits of all solar neutrino rates
together with the recoil electron energy spectrum have a confidence
level ranging between $\sim 10^{-3}$ (scenario with $\lam$ couplings)
and $\sim 10^{-4}$ (scenario with $\lam'$ couplings). The discrepancy
between the experimental results and the theoretical predictions is
due to the results from the SNO experiment.  The considered model,
which provides a realistic interpretation of the atmospheric neutrino
anomaly and respects the constraints on neutrino oscillations obtained
at reactor experiments (see Section \ref{approx}), does {\it not}
constitute an acceptable solution to the solar neutrino problem. This
analysis does not rely on the nature of the $\not\!\!R_p-$interactions
and thus applies to any new physics contributing to $R_{12}$ and
$R_{13}$, respecting the upper bound in Eq.~(\ref{eq:nineteen}). We
conclude that in this case a non-vanishing vacuum mixing angle for the
first generation neutrino is necessary. A future interesting
investigation would be to determine whether the same model, but based
on other types of lepton flavour violating interactions underlying the
Standard Model, can represent a reasonable solution to the atmospheric
and solar neutrino puzzle, or whether {\it all} such models can be
excluded.

\vspace{1cm}

\noindent {\bf \Large Acknowledgments}

\noindent We are grateful to V.~Barger and K.~Whisnant for interesting
discussions on their work \cite{Barger91}. We thank S.~Na\-kamura for
providing us with the latest results on the cross sections of the
charged current reaction. We also thank T.~Laval, W.~Mader, A.~Quadt
and G.~Wilquet, for their great help in using and interpreting the
results from the function minimization package {\tt MINUIT}.  G.~M.
acknowledges support from the {\it Alexander von Humboldt} Foundation,
the Belgian SSTC under contract IUAP and the French Community of
Belgium (ARC).



\begin{thebibliography}{99}

\bibitem{book} J.~N.~Bahcall,
{\it  CAMBRIDGE, UK: UNIV. PR. (1989), 567p}.
\bibitem{Barger80} V. Barger {\it et al.}, Phys. Rev. D {\bf 22} (1980) 2718.
\bibitem{MS} S. P. Mikheyev and A. Yu. Smirnov, Yad. Fiz. {\bf 42} 
    (1985) 1441 [Sov. J. Nucl. Phys. {\bf 42} (1985) 913]; 
Nuovo Cimento  {\bf 9C} (1986) 17.
\bibitem{Bethe86} H. A. Bethe, Phys. Rev. Lett. {\bf 56} (1986) 1305.
\bibitem{Rosen86} S. P. Rosen and J. M. Gelb, Phys. Rev. D {\bf 34} 
    (1986) 969.
\bibitem{Parke86} S. J. Parke, Phys. Rev. Lett. {\bf 57} (1986) 1275. 
\bibitem{rev1} S. M. Bilenky and S. T. Petcov, Rev. Mod. Phys. {\bf 59} 
    (1987) 671. 
\bibitem{rev2} T. K. Kuo and J. Pantaleone, Rev. Mod. Phys. {\bf 61} 
    (1989) 937. 
\bibitem{W} L. Wolfenstein, Phys. Rev. D {\bf 17} (1978) 2369.
\bibitem{Guzzo91} M. M. Guzzo, A. Masiero, S.T. Petcov, Phys. Lett. B
  {\bf 260} (1991) 154.
\bibitem{Roulet91} E. Roulet, Phys. Rev. D {\bf 44} (1991) 935.
  
\bibitem{Barger91} V.~D.~Barger, R.~J.~Phillips and K.~Whisnant,
Phys.\ Rev.\ D {\bf 44} (1991) 1629.


\bibitem{Krastev97} P.~I.~Krastev and J.~N.~Bahcall,
arXiv:hep-ph/9703267.

\bibitem{Berg00} S.~Bergmann, M.~M.~Guzzo, P.~C.~de Holanda,
  P.~I.~Krastev and H.~Nunokawa,
Phys.\ Rev.\ D {\bf 62} (2000) 073001
[arXiv:hep-ph/0004049].


\bibitem{Gago} A.~M.~Gago, M.~M.~Guzzo, P.~C.~de Holanda, H.~Nunokawa,
  O.~L.~Peres, V.~Pleitez and R.~Zukanovich Funchal,
  Phys.\ Rev.\ D {\bf 65} (2002) 073012 [arXiv:hep-ph/0112060].

  
\bibitem{Guzzo00} M.~M.~Guzzo, H.~Nunokawa, P.~C.~de Holanda and O.~L.~Peres,
Phys.\ Rev.\ D {\bf 64} (2001) 097301
[arXiv:hep-ph/0012089].

  
\bibitem{Berg97} S.~Bergmann,
Nucl.\ Phys.\ B {\bf 515} (1998) 363
[arXiv:hep-ph/9707398].


\bibitem{Valle01} M.~Guzzo, et al., 
Nucl.\ Phys.\ B {\bf 629} (2002) 479
[arXiv:hep-ph/0112310].

 
  
\bibitem{Valle02} J.~W.~Valle,
arXiv:hep-ph/0205216.

  
\bibitem{pre} R.~Adhikari, A.~Sil and A.~Raychaudhuri,
Eur.\ Phys.\ J.\ C {\bf 25} (2002) 125
[arXiv:hep-ph/0105119].



  
\bibitem{Berg99} S.~Bergmann, Y.~Grossman and D.~M.~Pierce,
Phys.\ Rev.\ D {\bf 61} (2000) 053005
[arXiv:hep-ph/9909390].

  
\bibitem{reac1} M.~Apollonio {\it et al.}  (CHOOZ Collab.),
Phys.\ Lett.\ B {\bf 466} (1999) 415
[arXiv:hep-ex/9907037].

  
\bibitem{reac2} 
F.~Boehm {\it et al.},
Phys.\ Rev.\ D {\bf 64} (2001) 112001
[arXiv:hep-ex/0107009].

  
\bibitem{reac3} V.~D.~Barger,
arXiv:hep-ph/0102052.

\bibitem{Salam} A.~Salam and J.~Strathdee,
Nucl.\ Phys.\ B {\bf 87} (1975) 85.



  
\bibitem{Fayet1} P.~Fayet,
Nucl.\ Phys.\ B {\bf 90} (1975) 104.



  
\bibitem{Haber} For reviews see: H. E. Haber and G. L. Kane, Phys.
  Rep.  {\bf 117} (1985) 75; H. P. Nilles, Phys. Rep. {\bf 110} (1984)
  1; and S.~P.~Martin, arXiv:hep-ph/9709356.
  
\bibitem{BP2000} J. N. Bahcall {\it et al.}, Astrophys. J. {\bf 555}
  (2001) 990; see also the J. N. Bahcall Web page:
  {\tt http://www.sns.ias.edu/$\sim$jnb/}

\bibitem{S17a} A.~R.~Junghans {\it et al.},
Phys.\ Rev.\ Lett.\  {\bf 88} (2002) 041101
[arXiv:nucl-ex/0111014].

  
\bibitem{S17b} J. N. Bahcall {\it et al.}, JHEP {\bf 0204} (2002) 007.

\bibitem{SNOa} Q. R. Ahmad {\it et al.} (SNO Collab.), Phys.
  Rev. Lett. {\bf 87} (2001) 071301.  

\bibitem{SNOb} Q. R. Ahmad {\it
    et al.} (SNO Collab.), Phys.  Rev. Lett. {\bf 89} (2002)
  011301.  

\bibitem{SNOc} Q. R. Ahmad {\it et al.} (SNO
  Collab.), Phys. Rev. Lett. {\bf 89} (2002) 011302.

\bibitem{SNOd} G.~A.~McGregor  (For the SNO Collaboration),
  arXiv:nucl-ex/0205006.


\bibitem{Drein} H.~K.~Dreiner,
arXiv:hep-ph/9707435.


\bibitem{Bhatt} G.~Bhattacharyya,
Nucl.\ Phys.\ Proc.\ Suppl.\  {\bf 52A} (1997) 83
[arXiv:hep-ph/9608415].

\bibitem{product} R. Barbier {\it et al.}, Report of the Group on
  R-parity violation (GDR-SUSY), arXiv:hep-ph/9810232.

\bibitem{actualize} F. Ledroit and G. Sajot, GDR-S-008 (1998), see:
  \vspace{-0.5cm}
\begin{verbatim}
http://qcd.th.u-psud.fr/GDR_SUSY/GDR_SUSY_PUBLIC/entete_note_publique
\end{verbatim}

\bibitem{perturb} B.~C.~Allanach, A.~Dedes and H.~K.~Dreiner,
Phys.\ Rev.\ D {\bf 60} (1999) 075014
[arXiv:hep-ph/9906209].



\bibitem{pubatm} Y. Fukuda {\it et al.} (Super-Kamiokande Collab.), 
    Phys. Rev. Lett. {\bf 81} (1998) 1562.

\bibitem{lastatm} T.~Toshito  (For the Super-Kamiokande Collaboration),
arXiv:hep-ex/0105023.


\bibitem{Haxton86} W. C. Haxton, Phys. Rev. Lett. {\bf 57} (1986) 1271.

\bibitem{Haxton87} W.~C.~Haxton,
Phys.\ Rev.\ D {\bf 35} (1987) 2352.


\bibitem{LZS} L.~D.~Landau, Phys. Z. Sowjetunion {\bf 2} (1932) 46;
    C.~Zener, Proc. R. Soc. London A {\bf 137} (1932) 696; 
    E.~C.~G.~Stueckelberg, Helv. Phys. Acta {\bf 5} (1932) 369.

\bibitem{Homestake} B. T. Cleveland {\it et al.}, Astrophys. J. {\bf 496} 
    (1998) 505; Nucl. Phys. (Proc. Suppl.) B {\bf 38} (1995) 47.

\bibitem{GALLEX} W. Hampel {\it et al.} (GALLEX Collab.), Phys.  
    Lett. B {\bf 447} (1999) 127. 

\bibitem{SAGE} J. N. Abdurashitov {\it et al.} (SAGE 
    Collab.), Phys. Rev. C {\bf 60} (1999) 055801.

\bibitem{SAGEbis} J. N. Abdurashitov {\it et al.} (SAGE 
    Collab.), arXiv:astro-ph/0204245.

\bibitem{Kamiokande} Y. Fukuda {\it et al.} (Kamiokande Collab.), 
    Phys. Rev. Lett. {\bf 77} (1996) 1683.

\bibitem{SuperK} S. Fukuda {\it et al.} (Super-Kamiokande Collab.), 
    Phys. Rev. Lett. {\bf 86} (2001) 5651.

\bibitem{XsecCl} J. N. Bahcall {\it et al.}, 
    Phys. Rev. C {\bf 54} (1996) 411.

\bibitem{XsecGm} J. N. Bahcall, 
    Phys. Rev. C {\bf 56} (1997) 3391.

\bibitem{XsecCC} S. Nakamura {\it et al.}, Nucl. Phys. A 
    {\bf 707} (2002) 561.

\bibitem{Lisi97} J. N. Bahcall {\it et al.}, 
    Phys. Rev. C {\bf 55} (1997) 494.

\bibitem{Eres1} K.~S.~Hirata {\it et al.},
Phys.\ Rev.\ D {\bf 38} (1988) 448.


\bibitem{Eres2} K. S. Hirata {\it et al.} (Kamiokande Collab.),
    Phys. Rev. Lett. {\bf 63} (1989) 16.

\bibitem{Eres3} K. S. Hirata {\it et al.} (Kamiokande Collab.),
    Phys. Rev. Lett. {\bf 65} (1990) 1297.

\bibitem{SuperK98} Y. Fukuda {\it et al.} (Super-Kamiokande 
Collab.), Phys. Rev. Lett. {\bf 82} (1999) 2430.

\bibitem{Nakahata99} M. Nakahata {\it et al.} (Super-Kamiokande 
Collab.), Nucl. Instrum. Methods Phys. Res. Sect. A {\bf 421} (1999) 113.

\bibitem{Faid97} B.~Faid, G.~L.~Fogli, E.~Lisi and D.~Montanino,
Phys.\ Rev.\ D {\bf 55} (1997) 1353
[arXiv:hep-ph/9608311].


\bibitem{Smirnov98} J.~N.~Bahcall, P.~I.~Krastev and A.~Y.~Smirnov,
Phys.\ Rev.\ D {\bf 58} (1998) 096016
[arXiv:hep-ph/9807216].


\bibitem{Garay99} M. C. Gonzalez-Garcia {\it et al.}, Nucl. Phys. B {\bf 573} 
    (2000) 3. 

\bibitem{Fogli95} G. L. Fogli and E. Lisi, Astropart. Phys. 
    {\bf 3} (1995) 185.

\bibitem{Guzzo98} M. M. Guzzo and H. Nunokawa, 
    Astropart. Phys. {\bf 12} (1999) 87.

\bibitem{BU} J. N. Bahcall and R. Ulrich, Rev. Mod. Phys. 
    {\bf 60} (1988) 297.

\bibitem{BP1992} J. N. Bahcall and M. H. Pinsonneault, Rev. Mod. Phys. 
    {\bf 64} (1992) 885. 

\bibitem{BP1995} J. N. Bahcall and M. H. Pinsonneault, Rev. Mod. Phys. 
    {\bf 67} (1995) 781.

\bibitem{BP1998} J. N. Bahcall {\it et al.}, Phys. Lett. B 
    {\bf 433} (1998) 1.

\bibitem{XsecCCpre} S. Nakamura {\it et al.}, 
    Phys. Rev. C {\bf 63} (2001) 034617.

\bibitem{dreiner} H.~K.~Dreiner and G.~G.~Ross,
Nucl.\ Phys.\ B {\bf 365} (1991) 597.

\bibitem{Barger89} V.~D.~Barger, G.~F.~Giudice and T.~Han,
Phys.\ Rev.\ D {\bf 40} (1989) 2987.


\bibitem{Roy96} D. Choudhury and P. Roy, Phys. Lett. B {\bf 378} 
    (1996) 153.

\bibitem{Mlim} ALEPH Collaboration, 
    {\it ``Search for Supersymmetric Particles with 
    R-Parity Violating Decays in $e^+ e^-$ Collisions at sqrt(s)
    up to $209 GeV$''}, ALEPH/2002-016,
    CONF/2002-005, ABS276, Contributed paper to ICHEP 2002.

\bibitem{Ellis95} J. Ellis {\it et al.}, Mod. Phys. Lett. A {\bf 10} 
    (1995) 1583.

\bibitem{Mohapatra86} R.~N.~Mohapatra,
Phys.\ Rev.\ D {\bf 34} (1986) 3457.


\bibitem{Hirsch95} M. Hirsch {\it et al.}, Phys. Rev. Lett. {\bf 75} (1995) 17;
    Phys. Rev. D {\bf 53} (1996) 1329.

\bibitem{GMSB} G. F. Giudice and R. Rattazzi, Phys. Rep. {\bf 322} 
    (1999) 421.

\bibitem{Kim97} J.~E.~Kim, P.~Ko and D.~G.~Lee,
Phys.\ Rev.\ D {\bf 56} (1997) 100
[arXiv:hep-ph/9701381].


\end{thebibliography}
\end{document}